\documentclass[conference]{IEEEtran}

\usepackage{amsfonts}
\usepackage{amssymb}
\usepackage{amsthm}
\usepackage{amsmath,amsfonts,amssymb}
\usepackage[dvips]{graphicx}
\usepackage{subfigure}
\usepackage{verbatim}
\usepackage{setspace}
\usepackage{bm}
\usepackage{algorithmic} 
\usepackage[ruled,vlined]{algorithm2e}
\usepackage{cite}

\usepackage{changepage}
\usepackage{pdfpages}
\usepackage{color}
\allowdisplaybreaks

\IEEEoverridecommandlockouts

\begin{document}
\title{\huge{Multiuser Communications with Movable-Antenna Base Station Via Antenna Position Optimization }}

\author{\IEEEauthorblockN{Xiangyu Pi$^\dag$, Lipeng Zhu$^\ddag$, Zhenyu Xiao$^\dag$, and Rui Zhang$^{\ddag\S}$}
	\IEEEauthorblockA{$^\dag$School of Electronic and Information Engineering, Beihang University, Beijing, China 100191.\\
		$^\ddag$Department of Electrical and Computer Engineering, National University of Singapore, Singapore 117583.\\
		$^\S$Shenzhen Research Institute of Big Data, The Chinese University of Hong Kong, Shenzhen, China 518172.\\
		E-mail: pixiangyu@buaa.edu.cn, zhulp@nus.edu.sg, xiaozy@buaa.edu.cn, elezhang@nus.edu.sg}
	\vspace{-0.58 cm}
}

\maketitle


\begin{abstract} 
	 This paper studies the deployment of multiple movable antennas (MAs) at the base station (BS) for enhancing the multiuser communication performance. 
	 First, we model the multiuser channel in the uplink to characterize the wireless channel variation caused by MAs' movement at the BS. Then, an optimization problem is formulated to maximize the minimum achievable rate among multiple users for MA-aided uplink multiuser communications by jointly optimizing the MAs' positions, their receive combining at the BS, and the transmit power of users, under the constraints of finite moving region of MAs, minimum inter-MA distance, and maximum transmit power of each user. 
	 To solve this challenging non-convex optimization problem, a two-loop iterative algorithm is proposed by leveraging the particle swarm optimization (PSO) method. Specifically, the outer-loop updates the positions of a set of particles, where each particle's position represents one realization of the antenna positioning vector (APV) of all MAs. The inner-loop implements the fitness evaluation for each particle in terms of the max-min achievable rate of multiple users with its corresponding APV, where the receive combining matrix of the BS and the transmit power of each user are optimized by applying the block coordinate descent (BCD) technique. Simulation results show that the antenna position optimization for MAs-aided BS can significantly improve the rate performance as compared to conventional BS with fixed-position antennas (FPAs). 
\end{abstract}

%
\IEEEpeerreviewmaketitle

\section{Introduction}
\IEEEPARstart{W}{ith} the development of sixth-generation (6G) and beyond wireless communication systems, there is an urgent need for exploring large-capacity and high-reliability communication technologies~\cite{Ma20216G,Wang20236G}. To achieve this goal, multi-user/multi-antenna or so-called multiple-input multiple-output (MIMO) communication technologies have been widely investigated to improve the spectral and energy efficiency by exploiting the spatial multiplexing~\cite{HBF2017MA,MIMO2021LQ}.
However, the antennas in conventional MIMO systems are deployed at fixed positions,  which cannot fully exploit the degrees of freedom (DoFs) in the continuous spatial domain for optimizing the spatial multiplexing performance. 

In order to overcome this fundamental limitation, movable antenna (MA) has been recently proposed as a new solution for fully exploiting the wireless channel variation in the continuous spatial domain~\cite{zhu2023movable,zhu2022modeling,ma2022mimo,MUMA2023ZL}. Different from conventional fixed-position antennas (FPAs), each MA is connected to the radio frequency (RF) chain via a flexible cable, which allows its position to be flexibly adjusted in a given spatial region with the aid of a diver component or by other means, for achieving more favorable channels to enhance the communication performance.
In~\cite{zhu2023movable}, the hardware architecture and channel characterization for MA systems were presented, and the advantages of MAs over conventional FPAs were demonstrated in terms of signal power improvement,
interference mitigation, flexible beamforming, and spatial multiplexing.
In~\cite{zhu2022modeling}, a field-response based channel model for MA-aided communication systems was developed, which characterizes the channel variation with respect to MAs' positions. Based on the field-response based channel model, the channel capacity of the MA-aided MIMO system was maximized in~\cite{ma2022mimo} by simultaneously adjusting the MAs' positions in transmitter and receiver located regions. Moreover, it was validated that jointly designing the positions of transmit and receive MAs can improve the multiplexing performance of MIMO systems. Besides, the total transmit power of multiple users was minimized in~\cite{MUMA2023ZL} by jointly optimizing the single-MA position and the transmit power of users, as well as the receive combining matrix of FPAs at the base station (BS) under the uplink multiuser communication setting. It was demonstrated that the MA-aided multiuser system can not only increase the channel gain but also achieve more effective interference mitigation over FPAs.
Note that the above performance improvement of MA systems relies on the availability of complete channel state information (CSI) between the entire transmit and receive regions where the antennas are located. 
To this end,
a  novel successive transmitter-receiver
compressed sensing (STRCS) method was proposed in~\cite{ma2023compressed} for channel estimation in  MA-aided communication systems, where the complete CSI between the transmit and receive regions was reconstructed by estimating the multi-path components.
However, none of the existing works~\cite{zhu2023movable,zhu2022modeling,ma2022mimo,MUMA2023ZL,ma2023compressed} investigated the employment of MAs at the BS to enhance the communication performance of multiple users.

In light of the above, this paper investigates a new MA-aided uplink multiuser communication system where multiple MAs are deployed at the BS to serve multiple users simultaneously, and each user is equipped with a single FPA. In particular, we study the joint optimization of MAs positioning, receive combining at the BS, and transmit power of different users to maximize their minimum achievable rate, subject to the constraints of finite moving region of MAs, minimum inter-MA distance, and maximum transmit power of each user.
To solve the formulated non-convex optimization problem with highly-coupled variables, a two-loop iterative algorithm is developed based on particle swarm optimization (PSO) to obtain a sub-optimal solution efficiently. 
In the inner-loop, for a given MAs positioning solution, transmit power of each user and receive combining matrix at the BS are jointly designed by applying the block coordinate descent (BCD) technique. In the outer-loop, a PSO-based algorithm is proposed to optimize MAs' positions, where the fitness function of each particle which represents a MAs positioning solution is the max-min achievable rate obtained in the inner-loop.
Simulation results demonstrate that compared to conventional BS with FPAs, the proposed new BS architecture with MAs can significantly improve the rate performance by antenna position optimization.

\section{System Model and Problem Formulation}
\begin{figure}[t]
	\begin{center}
		\includegraphics[width=8.8 cm]{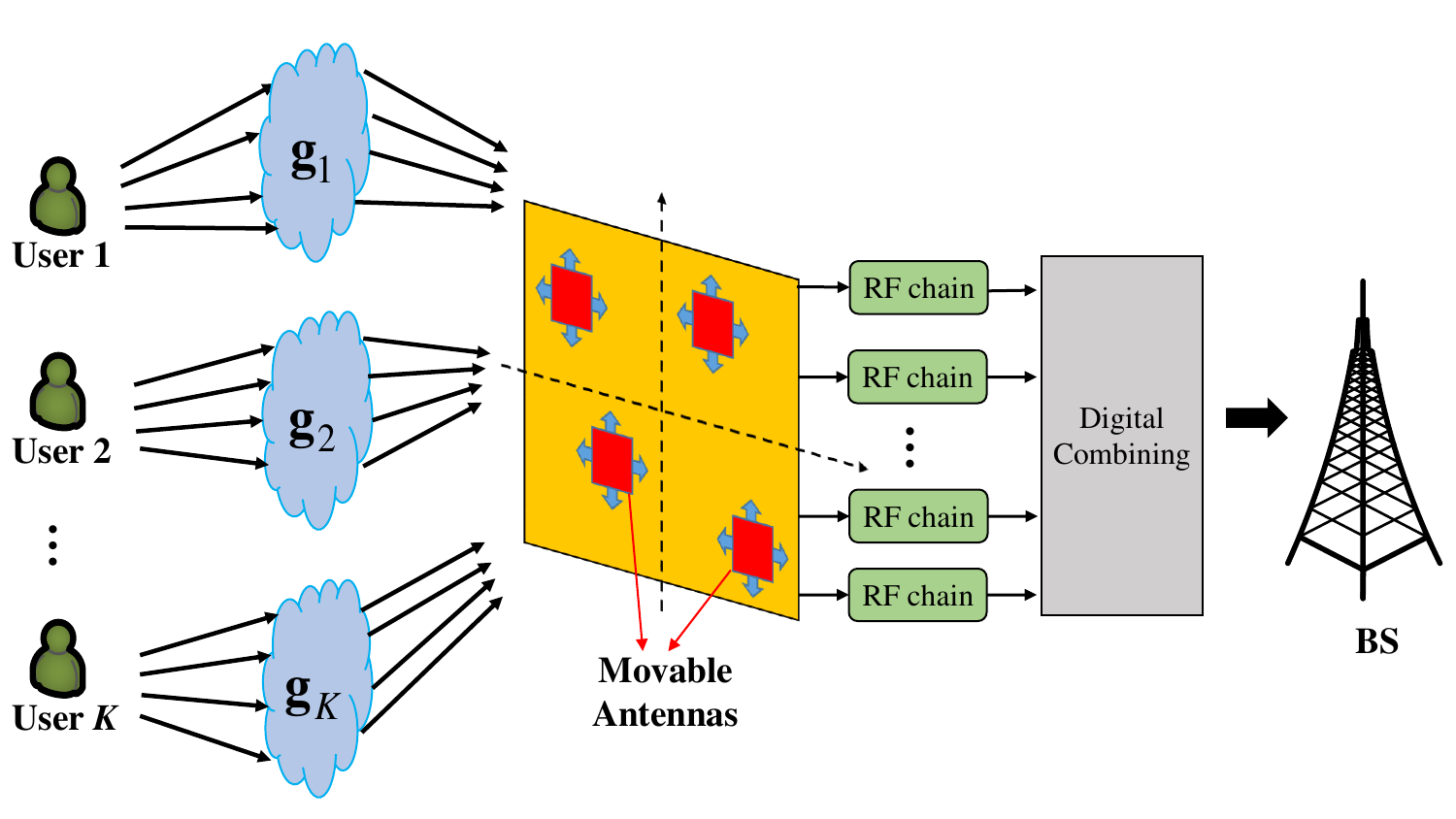}
		\caption{Illustration of the uplink transmission between $K$ FPA-users  and the BS equipped with $M$ MAs.}
		\label{fig:system}
	\end{center}
\vspace{-0.2cm}
\end{figure}

As shown in Fig.~\ref{fig:system}, $K$ single-FPA users are served by the BS equipped with $M$ MAs, each of which is connected to an RF chain via a flexible cable, thus enabled to move in a local two-dimensional (2D) region $\mathcal{C}_{r}$ at the BS for improving the channel conditions with users. We consider the space-division multiple access (SDMA) of users communicating with the BS in the uplink simultaneously, and thus the number of users is assumed not exceeding that of MAs at the BS, i.e., $K\le M$. The position of the $m$-th receive MA can be represented by its Cartesian coordinates, $\mathbf{r}_m=[x_m,y_m]^\mathrm{T} \in \mathcal{C}_{r}$ for $  1 \leq m \leq M$. Without loss of generality, the 2D region for antenna moving, i.e., $ \mathcal{C}_{r}$, is assumed as a square region of size $A\times A$. 

The received signal at the BS is processed using a digital combining matrix, which yields
\begin{equation}
	\label{signal_model}
	\mathbf{y}=\mathbf{W}^{\mathrm{H}} \mathbf{H}(\tilde{\mathbf{r}}) \mathbf{P}^{1 / 2} \mathbf{s}+\mathbf{W}^{\mathrm{H}} \mathbf{n},
\end{equation}
where $\mathbf{W} =\left[\mathbf w_1,\mathbf w_2,\cdots,\mathbf w_K\right]\in\mathbb C^{M\times K}$  is the receive combining matrix at the BS, $\mathbf{H}(\tilde{\mathbf{r}})=[\mathbf{h}_1(\tilde{\mathbf{r}}),\mathbf{h}_2(\tilde{\mathbf{r}}),\cdots,\mathbf{h}_K(\tilde{\mathbf{r}})]\in\mathbb{C}^{M\times K}$ is the channel matrix from all $K$ users to the $M$ MAs at the BS with $\tilde{\mathbf{r}}=[\mathbf{r}^\mathrm{T}_1,\mathbf{r}^\mathrm{T}_2,\cdots,\mathbf{r}^\mathrm{T}_M]^\mathrm{T}$ denoting the antenna positioning
vector (APV) for MAs, $\mathbf{P}^{1/2}=\mathrm{diag}\left\{\sqrt p_1,\sqrt p_2,\cdots,\sqrt {p_K}\right\}$ is the power matrix with $p_k , 1 \leq k \leq K$, representing the transmit power of user $k$, $\mathbf{s}$ is the independent
and identically distributed (i.i.d.) transmit signal vector of users each with normalized power, i.e., $\mathbb{E}(\mathbf{s}\mathbf{s}^{\mathrm{H}})=\mathbf{I}_K$, and $\mathbf{n}\sim \mathcal{CN}(0,\sigma^2\mathbf{I}_M)$ is the zero-mean additive white Gaussian noise (AWGN) with average power $\sigma^2$.
\subsection{Channel Model}
We employ the filed-response based channel model in~\cite{zhu2022modeling}, where the channel response is the superposition of the coefficients of multiple channel paths between the transceivers. 
Let $L_{k}$ denote the total number of receive channel paths at the BS from user $k$, $1 \leq k \leq K$. Then, the signal propagation phase difference of the $l$-th path  for user $k$ between the position of the $m$-th MA and the reference point at the BS, $\mathbf{r}_{0}=[0, 0]^T$, is written as 
\begin{equation}
	\label{phase difference}
	\rho_{k,l}(\mathbf{r}_m)=x_m \sin{\theta_{k,l}}\cos{\phi_{k,l}}+y_m \cos{\theta_{k,l}},
\end{equation}
where $\theta_{k,l}$ and $\phi_{k,l}$  are the elevation and azimuth AoAs for the $l$-th receive path between the user $k$ and the BS. Accordingly, the field-response vector (FRV) of the receive channel paths between the user $k$ and the $m$-th MA at the BS is given by ~\cite{zhu2022modeling}
\begin{equation}
	\label{FRV}
	\mathbf{f}_k(\mathbf{r}_m)=\left[e^{j\frac{2\pi}{\lambda}\rho_{k,1}(\mathbf{r}_m)},e^{j\frac{2\pi}{\lambda}\rho_{k,2}(\mathbf{r}_m)},\dots,e^{j\frac{2\pi}{\lambda}\rho_{k,L_k}(\mathbf{r}_m)}\right]^\mathrm{T}.
\end{equation}
As such, the channel vector between user $k$ and the BS is obtained as 
\begin{equation}
	\label{channel_model}
	\mathbf{h}_k(\tilde{\mathbf{r}})=\mathbf{F}_k^{\mathrm{H}}(\tilde{\mathbf{r}})\mathbf{g}_k,
\end{equation}
where $\mathbf{F}_k(\tilde{\mathbf{r}})=[\mathbf{f}_k(\mathbf{r}_1),\mathbf{f}_k(\mathbf{r}_2),\cdots,\mathbf{f}_k(\mathbf{r}_M)]\in\mathbb{C}^{L_k\times M}$ donates the field-response matrix (FRM) at the BS,
and $\mathbf{g}_k=[g_{k,1},g_{k,2},\cdots,g_{k,L_k}]^{\mathrm T}$ is the path-response vector (PRV), which represents the multi-path response coefficients from user $k$ to the reference point in the receive region. As can be observed, the channel coefficient $[\mathbf{h}_k(\tilde{\mathbf{r}})]_m=\mathbf{f}_k(\mathbf{r}_m)^{\mathrm H}\mathbf{g}_k$ between the $m$-th MA and user $k$ is the sum of all elements of $\mathbf{g}_k$ weighted by the unit-modulus elements in $\mathbf{f}_k(\mathbf{r}_m)^{\mathrm H}$.  
As a result, small movement of each MA can change the channel vectors of all users significantly due to the phase variations of multiple channel paths (while their amplitude variations are relatively much less and thus negligible).
\subsection{Problem Formulation}
At the BS, the receive signal-to-interference-plus-noise ratio (SINR) for user $k$ is given by
\begin{equation}
	\label{SINR}
	\gamma_k=\frac{\left|\mathbf{w}_k^{\mathrm{H}}\mathbf{h}_k(\tilde{\mathbf{r}})\right|^2p_k}
	{\sum\limits _{i=1,i\ne k}^{K}\left|\mathbf{w}^{\mathrm{H}}_k\mathbf{h}_i(\tilde{\mathbf{r}})\right|^2p_i+\left\|\mathbf{w}_{k}\right\|_2^2\sigma^2}.
\end{equation}
Thus, the achievable rate for user $k$ is calculated as
\begin{equation}
	\label{Rate}
	R_k=\log_{2}{\left (1+ \gamma_k\right)}.
\end{equation}

In this paper, we aim to maximize the minimum achievable rate among all users to improve the overall performance by jointly optimizing the APV of MAs at the BS, i.e., $\tilde{\mathbf{r}}$, their receive combining matrix, i.e., $\mathbf{W}$, and the transmit power matrix, i.e., $\mathbf{P}$.
The max-min rate optimization problem is formulated as\footnote{The field-response information in the angular domain, including AoAs and PRVs, is assumed to be known, which can be acquired by using channel estimation methods for MA systems, such as STRCS in ~\cite{ma2023compressed}.
}
\begin{subequations}\label{op}
	\begin{align}
		\max \limits _{\tilde{\mathbf{r}},\mathbf W,\mathbf P}&~~\min \limits _{k}~~\{R_k\} \label{opA}\\
		\mbox { s.t.}~~ 
		& {\mathbf{r}}_m \in \mathcal{C}_{r}, 1 \le  m  \le M, \label{opB}\\
		& \left\|\mathbf{r}_{m}-\mathbf{r}_{i}\right\|_{2} \geq D, 1 \le  m \neq i \le M, \label{opC}\\
		& 0\le p_k \le p_{\mathrm{max}}, 1 \le k \le K.\label{opD}
	\end{align}   
\end{subequations}
Constraint (\ref{opB}) indicates that each MA can only move in the given receive region, $\mathcal{C}_{r}$. 
Constraint (\ref{opC}) ensures that minimum inter-MA distance $D$ at the BS for practical implementation. 
Constraint (\ref{opD}) ensures that transmit power of each user is non-negative and does
not exceed its maximum value, $p_{\mathrm{max}}$.
Note that problem (\ref{op}) is an non-convex optimization problem with highly coupled variables. Existing optimization tools cannot be directly used to obtain the globally optimal solution for problem (\ref{op}) with polynomial complexity in terms of $M$ and $K$.

\section{Proposed Solution}
Since there are three highly coupled matrices/vectors in the optimization variables of problem (\ref{op}), the conventional alternating optimization method approach by optimizing one of them with the other two being fixed  may not work well as it may lead to an undesired local optimal solution.  
To address this problem, we propose a two-loop iterative algorithm based on PSO. In the inner-loop, for any given APV, a BCD-based algorithm is developed to iteratively solve the receive combining and transmit power optimization. In the outer-loop, a PSO-based algorithm is applied to optimize the APV, where the fitness function of each particle (i.e., APV) is the max-min achievable rate obtained in the inner-loop.

\subsection{Receive Combining and Transmit Power Optimization}
In the inner-loop of the proposed algorithm, in order to calculate the fitness value of each particle, which represents an APV solution, we need to solve the following problem to determine the receive combining matrix and transmit power matrix for any given APV:  
\begin{subequations}\label{op:PandW}
	\begin{align}
		\max \limits _{\mathbf W,\mathbf P}&~~\min \limits _{k}~~\left\{R_k\right\} \label{op:PandW_A}\\
		\mbox { s.t.}~~
		&0\le p_k \le p_{\mathrm{max}}, 1 \le k \le K.\label{op:PandW_B}
	\end{align}
\end{subequations}

Note that for any given APV $\tilde{\mathbf r}$ and transmit power matrix $\mathbf P$, 
the optimal receive combining matrix $\mathbf W$ can be derived in closed form based on the minimum mean square error (MMSE) receiver~\cite{MMSE2022Lin}, i.e.,
\begin{equation}
	\label{MMSE}
	\begin{split}
		\hat{\mathbf W}(\tilde{\mathbf r},\mathbf P)&=\left(\mathbf H(\tilde{\mathbf r})\mathbf P\mathbf{H}(\tilde{\mathbf r})^{\mathrm H}+\sigma^2\mathbf I_M\right)^{-1}\mathbf {H}(\tilde{\mathbf r})\\
		&\triangleq\left[\hat{\mathbf{w}}_1,\hat{\mathbf{w}}_2,\cdots,\hat{\mathbf{w}}_K\right],\\   
	\end{split}
\end{equation}
with $\hat{\mathbf{w}}_k=\left(\mathbf H(\tilde{\mathbf r})\mathbf P\mathbf{H}(\tilde{\mathbf r})^{\mathrm H}+\sigma^2\mathbf I_M\right)^{-1}\mathbf{h}_k(\tilde{\mathbf r})$. Substituting (\ref{MMSE}) into (\ref{SINR}), the receive SINR for user $k$ given in (\ref{SINR}) can be rewritten as
\begin{equation}
	\label{SINR_OPpower}
	\hat{\gamma}_k=\frac{p_k[\mathbf{A}]_{k,k}}
	{\sum\limits _{i=1,i\ne k}^{K}p_i[\mathbf{A}]_{k,i}+b_k},
\end{equation}
where $[\mathbf{A}]_{k,i}= \left|\hat{\mathbf{w}}^{\mathrm{H}}_k\mathbf{h}_i(\tilde{\mathbf{r}})\right|^2, 1 \leq k,i \leq K, $ is the entry in the $k$-th row and $i$-th column of matrix $\mathbf{A}\in\mathbb{C}^{K\times K}$ and $b_k = \left\|\hat{\mathbf{w}}_{k}\right\|_2^2\sigma^2, 1 \leq k \leq K, $ is the $k$-th entry of column vector $\mathbf{b}=[b_1,b_2,\cdots,b_K]^\mathrm{T}\in\mathbb{C}^{K\times 1}$.

For any given APV $\tilde{\mathbf r}$ and receive combining matrix $\mathbf W$, in order to distinguish with the transmit power matrix used to calculate the receive combining matrix in the previous iteration, we introduce the transmit power vector $\mathbf{p}=[p_1,p_2,\cdots,p_k]^{\mathrm{T}}$ as an intermediate variable in the current iteration.
Therefore,
problem (\ref{op:PandW}) can be equivalently transformed into
\begin{subequations}\label{op_power}
	\begin{align}
		\max \limits _{\mathbf p,\eta}&~~\eta \label{op_powerA}\\
		\mbox { s.t.}~~ 
		& \hat{\gamma}_k \ge \eta, 1 \le k \le K,\label{op_powerB}\\
		& 0\le p_k \le p_{\mathrm{max}}, 1 \le k \le K,\label{op_powerC}
	\end{align}   
\end{subequations}
where $\eta$ represents the minimum SINR among the users. It is easy to verify that the optimal solution for problem (\ref{op_power}) is obtained as the constraints in (\ref{op_powerB}) are met with equality \cite{Maxmin2008Nace}. Otherwise, we can always adjust the transmit power of certain users to ensure the equality holds with the minimum SINR unchanged.
In other words, the linear equations $p_k[\mathbf{A}]_{k,k}/\eta=\sum\limits _{i=1,i\ne k}^{K}p_i[\mathbf{A}]_{k,i}+b_k,1\le k \le K,$ always hold,
which is equivalent to the following matrix form of linear equations with respect to $\mathbf{p}$:
\begin{equation}
	\label{Linear_equation}
	\mathbf{D(\eta)} \mathbf{p} = \mathbf b,
\end{equation}
where $\mathbf{D(\eta)}\in\mathbb C^{K \times K}$ is a square matrix whose diagonal elements and non-diagonal elements are given by $[\mathbf{D}(\eta)]_{k,k}=[\mathbf{A}]_{k,k}/\eta$ and $[\mathbf{D}(\eta)]_{k,i}=-[\mathbf{A}]_{k,i}$ for $1 \le k \ne i \le K$, respectively.
Thus, the transmit power vector can be expressed as a function with respect to $\eta$ as
\begin{equation}
	\label{power}
	{\mathbf{p}(\eta)} = \mathbf{D(\eta)}^{-1} \mathbf b.
\end{equation}

It is worth emphasizing that the solution for transmit power vector shown in  (\ref{power}) is feasible to problem (\ref{op_power}) only if constraint (\ref{op_powerC}) is satisfied. Thus, we develop the bisection method to find the maximum $\eta$ which makes ${\mathbf{p}(\eta)}$ satisfy constraint (\ref{op_powerC}). First, we choose an initial search interval $(\eta_{\mathrm{min}},\eta_{\mathrm{max}})$ 
with $\eta_{\mathrm{min}}=0$ and $\eta_{\mathrm{max}}=p_{\mathrm{max}}h_{\mathrm{min}}/\sigma^2$, where $h_{\mathrm{min}}$ is the minimum channel gain among users, i.e.,
$h_{\mathrm{min}}=\mathrm{Min}\left\{ \left\|\mathbf{h}_{1}(\tilde{\mathbf{r}})\right\|_2^2,\left\|\mathbf{h}_{2}(\tilde{\mathbf{r}})\right\|_2^2,\cdots,\left\|\mathbf{h}_{k}(\tilde{\mathbf{r}})\right\|_2^2\right\}$.
Then, the feasibility of the middle point of search internal,  $\eta=(\eta_{\mathrm{min}}+\eta_{\mathrm{max}})/{2}$, is examined by checking
whether $\mathbf{p}(\eta)$ satisfies constraint (\ref{op_powerC}). If $\eta$ is feasible, we update $\eta_{\mathrm{min}}$ as $\eta$, and otherwise update $\eta_{\mathrm{max}}$ as $\eta$. This process is repeated until $\eta_{\mathrm{max}} - \eta_{\mathrm{min}} <\epsilon$ where $\epsilon$ is a positive convergence threshold.

Based on the above analysis, a BCD-based algorithm is developed to jointly optimize the receive combining matrix and transmit power matrix with given APV $\tilde{\mathbf r}$. In each iteration, for given transmit power matrix $\mathbf{P}$, we obtain a closed-form solution for receive combining matrix $\mathbf{W}$ according to (\ref{MMSE}). For given $\mathbf{W}$, we then solve the transmit power vector $\mathbf{p}$ by using bisection method, and update it into $\mathbf{P}=\mathrm{diag}\{\mathbf{p}\}$. 
During the iterations, the receive combining matrix and transmit power matrix are alternately optimized until the increase on the
objective value in (\ref{op:PandW_A}) is below a small positive value $\xi$. 
The detailed BCD-based algorithm is shown in Algorithm~\ref{BCD}.
In line 1, the transmit power of all users is initialized to the maximum power, i.e., $\mathbf{P}^{(0)}=p_{\mathrm{max}}\mathbf{I}_K$. Then the channel matrix $\mathbf{H}(\tilde{\mathbf r})$ is calculated in line 2. With the input $\mathbf{P}^{(0)}$ and $\mathbf{H}(\tilde{\mathbf r})$, the initial receive combining matrix $\mathbf W^{(0)}$ is obtained by the MMSE receiver in line 3. Subsequently, the receive combining matrix and transmit power vector are alternately optimized in lines 4-10 until convergence. Note that in line 8, the minimum achievable rate among multiple users in the $j$-th iteration is defined as
\begin{equation}
	\label{RATE_PandV}
	\mathcal{G}\left(\mathbf{P}^{(j)},\mathbf W^{(j)}\right) = \operatorname*{min}_{k}~\left\{R_k\right\},
\end{equation}
where $R_k$ can be calculated by (\ref{SINR}) and (\ref{Rate}) with given $\mathbf{P}^{(j)}$ and $\mathbf W^{(j)}$. The iteration process will terminate if the relative increase of the objective value is below a convergence threshold $\xi$. Finally, the optimal receive combining matrix and transmit power matrix are obtained, which corresponds to the max-min achievable rate of multiple users for the given APV, i.e., $R(\tilde{\mathbf r}) =\mathcal{G}\left(\mathbf{P},\mathbf W\right)$.

\begin{algorithm}[t] \small
	\label{BCD}
	\caption{BCD-based algorithm for solving problem~(\ref{op:PandW}).}
	\begin{algorithmic}[1]
		\REQUIRE ~$\tilde{\mathbf r}$, $M$, $K$, $\lambda$, $p_{\mathrm{max}}$, $\sigma^2$, $\{g_k\}$, $\{\theta_{k,l}\}$, $\{\phi_{k,l}\}$, $\epsilon$, $\xi$.
		\ENSURE ~$\mathbf W$, $\mathbf P$, $R(\tilde{\mathbf r})$. \\
		\STATE Set the iteration index as $j=1$, and initialize
		$\mathbf{P}^{(0)}=p_{\mathrm{max}}\mathbf{I}_K$.  
		\STATE Calculate the channel response matrix $\mathbf{H}(\tilde{\mathbf r})$ according to (\ref{channel_model}) for given $\tilde{\mathbf r}$.
		\STATE Initialize the receive combining matrix $\mathbf W^{(0)}$ according to (\ref{MMSE}) for given $\mathbf{P}^{(0)}$ and $\tilde{\mathbf r}$.
		\REPEAT
		\STATE Calculate the transmit power vector $\mathbf{p}^{(j)}$ via bisection method for given $\mathbf{W}^{(j-1)}$ and $\mathbf{H}(\tilde{\mathbf r})$.
		\STATE Update $\mathbf{P}^{(j)}=\text{diag}\{\mathbf{p}^{(j)}\}$.
		\STATE Calculate the receive combining matrix $\mathbf W^{(j)}$ according to (\ref{MMSE}) for given $\mathbf{P}^{(j)}$ and $\mathbf{H}(\tilde{\mathbf r})$.
		\STATE Calculate $\mathcal{G}\left(\mathbf{P}^{(j)},\mathbf W^{(j)}\right)=\operatorname*{min}\limits_{k}~\left\{R_k\right\}$ for given $\mathbf{P}^{(j)}$ \quad and $\mathbf{W}^{(j)}$.
		\STATE Update $j\leftarrow j+1$.
		\UNTIL{$\left|\mathcal{G}\left(\mathbf{P}^{(j)},\mathbf W^{(j)}\right)-\mathcal{G}\left(\mathbf{P}^{(j-1)},\mathbf W^{(j-1)}\right)\right|<\xi$}
		\STATE Set the transmit power matrix $\mathbf P$ as $\mathbf{P}^{(j)}$.
		\STATE Set the receive combining matrix $\mathbf W$ as $\mathbf W^{(j)}$ .
		\STATE Calculate the maximum objective value for given APV $\tilde{\mathbf r}$, $R(\tilde{\mathbf r}) =\mathcal{G}\left(\mathbf{P},\mathbf W\right)$.
		\RETURN  $\mathbf W$, $\mathbf P$, $R(\tilde{\mathbf r})$.
	\end{algorithmic}
\end{algorithm}

\subsection{APV Optimization}
In the outer-loop of proposed algorithm, since the optimal receive combining matrix and transmit power matrix for any given APV $\tilde{\mathbf{r}}$ can be calculated in the inner-loop, the corresponding max-min achievable rate for multiple users can be accordingly expressed as a function for the APV, i.e.,  $R(\tilde{\mathbf{r}})$. Thus, the original problem~(\ref{op}) can be transformed to the following APV optimization problem
\begin{subequations}\label{op_APV}
	\begin{align}
		\max \limits _{\tilde{\mathbf{r}}}&~~R(\tilde{\mathbf{r}}) \label{op_APVA}\\
		\mbox { s.t.}~~ 
		& {\mathbf{r}}_m \in \mathcal{C}_{r}, 1 \le  m  \le M \label{op_APVB}\\
		& \left\|\mathbf{r}_{m}-\mathbf{r}_{i}\right\|_{2} \geq D, 1 \le  m \neq i \le M. \label{op_APVC}
	\end{align}   
\end{subequations}

To solve this difficult problem, PSO is introduced as an efficient approach~\cite{PSO2019Zhu}.
In the PSO based algorithm, we first randomly initialize $N$ particles with positions $\mathcal{R}^{(0)}=\{\tilde{\mathbf{r}}_1^{(0)},\tilde{\mathbf{r}}_2^{(0)},...,\tilde{\mathbf{r}}_{N}^{(0)}\}$  and velocities  $\mathcal{V}^{(0)}=\{\tilde{\mathbf{v}}_1^{(0)},\tilde{\mathbf{v}}_2^{(0)},...,\tilde{\mathbf{v}}_{N}^{(0)}\}$, where each particle represents a possible solution for the APV.
Then, each particle updates its position according to the known local best position, i.e., $\tilde{\mathbf{r}}_{n,pbest}$ and the known global best position, i.e., $\tilde{\mathbf{r}}_{gbest}$.
Thus, for each iteration, the velocity and position of each particle are updated as
\begin{equation}
	\label{update_velocity}
	\tilde{\mathbf{v}}_{n}^{(t+1)}=\omega \tilde{\mathbf{v}}_{n}^{(t)} + c_1 \tau_1 \left(\tilde{\mathbf{r}}_{n,pbest}-\tilde{\mathbf{r}}_{n}^{(t)}\right)+c_2 \tau_2 \left(\tilde{\mathbf{r}}_{gbest}-\tilde{\mathbf{r}}_{n}^{(t)}\right),
\end{equation}
\begin{equation}
	\label{update_position}
	\tilde{\mathbf{r}}_{n}^{(t+1)}=\mathcal{B}\left( \tilde{\mathbf{r}}_{n}^{(t)} + \tilde{\mathbf{v}}_{n}^{(t+1)} \right),
\end{equation}
for $1 \le n \le N$ with $t$ representing the iteration index. Parameters $c_1$ and $c_2$ are the individual and global learning factors, which represent the step size of each particle moving toward the best position.  $\tau_1$ and $\tau_2$ are two random parameters uniformly distributed in $[0,1]$,  which aim to increase the randomness of the search  for escaping from local optima. $\omega$ is the inertia weight, which is used to maintain the inertia of the particle movement. 

Due to constraint (\ref{op_APVB}), if a particle moves out of the boundary of the feasible region, we project its position component to the corresponding minimum/maximum value, i.e., 
\begin{equation}
	\label{bound_p}
	[\mathcal{B}(\tilde{\mathbf{r}})]_i=\left\{\begin{array}{cc}
		-\frac{A}{2},&\text{if}~~[\tilde{\mathbf{r}}]_i<-\frac{A}{2},\\ 
		\frac{A}{2},&\text{if}~~[\tilde{\mathbf{r}}]_i>\frac{A}{2},\\ \left[\tilde{\mathbf{r}}\right]_{i},&\text{otherwise},\end{array}\right.
\end{equation}
where $[\tilde{\mathbf{r}}]_i$ denotes the $i$-th element of $\tilde{\mathbf{r}}$. The utilization of projection function $\mathcal{B}(\tilde{\mathbf{r}})$ in~(\ref{update_position}) is to ensure that the solution for APV is always located in the feasible region during the iterations.

The fitness of each particle is evaluated in the inner-loop and is given by $R(\tilde{\mathbf{r}}_{n}),1 \leq n \leq N$, for maximizing the minimum achievable rate of multiple users under the given APV. Moreover, in order to ensure constraint (\ref{op_APVC}), we introduce an adaptive penalty factor to the fitness function and update it as follows~\cite{DE2011DS}
\begin{equation}
	\label{penalty}
	\mathcal{F}\left(\tilde{\mathbf{r}}_{n}^{\left(t\right)}\right)=R\left(\tilde{\mathbf{r}}_{n}^{\left(t\right)}\right)-\tau \left|\mathcal{P}\left(\tilde{\mathbf{r}}_{n}^{\left(t\right)}\right)\right|,
\end{equation}
where $\mathcal{P}\left(\tilde{\mathbf{r}}\right)$ is a set with the cardinality $\left|\mathcal{P}\left(\tilde{\mathbf{r}}\right)\right|$, in which each entry represents a pair of MAs in the APV $\tilde{\mathbf{r}}$ that violate the minimum inter-MA distance constraint. It can be defined as
\begin{equation}
	\label{powerset}
	\mathcal{P}\left(\tilde{\mathbf{r}}\right)=\left\{ (\mathbf{r}_{m},\mathbf{r}_{i})| \left\|\mathbf{r}_{m}-\mathbf{r}_{i}\right\|_{2} < D, 1 \le  m < i \le M\right\}.
\end{equation}
$\tau$ is a large positive penalty parameter which ensures that the inequality equation  $R\left(\tilde{\mathbf{r}}_{n}^{\left(t\right)}\right)-\tau \le 0$ holds for all APVs. Thus, 
during the iteration process, the penalty factor enforces $\left|\mathcal{P}\left(\tilde{\mathbf{r}}_{n}^{\left(t\right)}\right)\right|$ to approach zero, i.e., constraint (\ref{op_APVC}) is satisfied eventually. 

With the fitness evaluation conducted on each particle, their individual and global best positions are updated until convergence. The final best position among the particles is generally a suboptimal solution for APV, and its corresponding  receive combining matrix and transmit power matrix are calculated by Algorithm~\ref{BCD}. The detailed PSO-based overall algorithm for solving problem (\ref{op}) is summarized in Algorithm~\ref{PSO}.
In line 1, in the $2M$-dimensional search space, the position and velocity of each particle are randomly initialized with each component uniformly distributed in $[-A/2,A/2]$. In lines 2-3, each particle is evaluated by the fitness function, thus finding the local and global best position. In line 6, the velocity of each particle is updated according to the relative local and global best positions, which drive the particle moving in the feasible region. In lines 7-13, we evaluate the particle’s fitness value and compare it with that of its local/global best position. 
For each particle, if its fitness value is better than that of its local best position or the global best position, then the corresponding best locations are replaced with the current particle’s position. 
Thus, in lines 4-15, the global best position can be updated with its fitness value non-decreasing during the iterations.

Hereto, we have solved the original problem~(\ref{op}). In the proposed solution,
the receive combining matrix and transmit power matrix are optimal,
while the APV is suboptimal in general .

\subsection{Convergence and Complexity Analysis}\label{sec_ConvergenceComplexity}
Since the overall algorithm is two-loop based, its convergence depends on the convergence of BCD-based algorithm in the inner-loop and PSO-based algorithm in the outer-loop. 
The convergence of  Algorithm \ref{BCD} is guaranteed by the following inequality:
\begin{equation}
	\label{convergence_inner}
	\begin{split}
		\mathcal{G}\left(\mathbf{P}^{(j)},\mathbf W^{(j)}\right)
		&=
		\mathcal{G}\left(\mathbf{P}^{(j)},\hat{\mathbf W}\left(\tilde{\mathbf r},\mathbf P^{(j)}\right)\right)\\
		&\overset{(a)}{\ge} 
		\mathcal{G}\left(\mathbf{P}^{(j)},\hat{\mathbf W}\left(\tilde{\mathbf r},\mathbf P^{(j-1)}\right)\right)\\
		&\overset{(b)}{\ge} 
		\mathcal{G}\left(\mathbf{P}^{(j-1)},\hat{\mathbf W}\left(\tilde{\mathbf r},\mathbf P^{(j-1)}\right)\right)\\
		&=
		\mathcal{G}\left(\mathbf{P}^{(j-1)},\mathbf W^{(j-1)}\right),
	\end{split}
\end{equation}
where $(a)$ holds because $\hat{\mathbf W}\left(\tilde{\mathbf r},\mathbf P^{(j)}\right)$ is the optimal MMSE combining matrix for maximizing the SINR of each user under the current transmit power $\mathbf{P}^{(j)}$, and $(b)$ holds since $\mathbf{P}^{(j)}$ is the optimal transmit power searched by bisection method under the current MMSE combining matrix $\hat{\mathbf W}\left(\tilde{\mathbf r},\mathbf P^{(j-1)}\right)$. It means that the objective value is non-decreasing during the iterations in Algorithm~\ref{BCD}.

Moreover, the fitness value of the global best position is non-decreasing during the iterations in Algorithm~\ref{PSO}, i.e.,
\begin{equation}
	\label{convergence_outter}	
	\mathcal{F}\left(\tilde{\mathbf r}_{gbest}^{\left(t+1\right)}\right)\ge \mathcal{F}\left(\tilde{\mathbf r}_{gbest}^{\left(t\right)}\right).
\end{equation} 
Meanwhile, the objective value of problem (\ref{op}) is always bounded. Thus, the convergence of the overall algorithm is guaranteed. Moreover, the convergence performance will also be validated by simulation in Section~~\ref{sec_SimulationResults}.


The computational complexity of bisection method for solving problem (\ref{op_power}) is 
$\mathcal{O}(K^3\log_{2}{\epsilon^{-1}} )$, depending on the search accuracy $\epsilon$ and the number of users $K$.
Denoting the maximum number of iterations of Algorithm~\ref{BCD} for 
solving problem (\ref{op:PandW}) as $J$, the corresponding computational complexity is given by $\mathcal{O}\left(J(M^3+K^3\log_{2}{\epsilon^{-1}} )\right)$.
As a result, with the swarm size $N$ and the maximum number of iterations $T$, the maximum computational complexity of Algorithm~\ref{PSO} for solving problem (\ref{op}) is $\mathcal{O}\left(NTJ(M^3+K^3\log_{2}{\epsilon^{-1}} )\right)$.


\begin{algorithm}[t] \small
	\label{PSO}
	\DontPrintSemicolon
	\SetAlgoLined
	\caption{PSO-based Algorithm for solving problem~(\ref{op}).}
	\begin{algorithmic}[1]
		\REQUIRE ~$M$, $K$, $\mathcal{C}_r$, $\lambda$, $p_{\mathrm{max}}$, $\sigma^2$, $\{\mathbf g_k\}$, $\{\theta_{k,l}\}$, $\{\phi_{k,l}\}$, $N$, $T$, $\epsilon$,\quad $\xi$, $c_1$, $c_2$, $\omega$, $\tau$.
		\ENSURE ~$\tilde{\mathbf r},\mathbf{W}, \mathbf{P}$. \\
		\STATE Initialize the $N$ particles with positions $\mathcal{R}^{(0)}$ and velocities $\mathcal{V}^{(0)}$.
		\STATE Evaluate the fitness value for each particle using Algorithm~\ref{BCD}.
		\STATE Obtain the local best position $\tilde{\mathbf{r}}_{n,pbsest}=\tilde{\mathbf{r}}_{n}^{\left(0\right)}$ for $1\le n \le N$ and the global best position $\tilde{\mathbf{r}}_{gbest}=\mathop{\arg\max}\limits _{\tilde{\mathbf{r}}_{n}^{\left(0\right)}}\{\mathcal{F}\left(\tilde{\mathbf{r}}_{1}^{\left(0\right)}\right),\mathcal{F}\left(\tilde{\mathbf{r}}_{2}^{\left(0\right)}\right),...,\mathcal{F}\left(\tilde{\mathbf{r}}_{N}^{\left(0\right)}\right)\}$.
		\FOR{$t=1$ to $T$}
		\FOR{$n=1$ to $N$}
		\STATE Update the velocity and position of particle $n$  according to (\ref{update_velocity}) and (\ref{update_position}), respectively.
		\STATE  Evaluate the fitness value of particle $n$ using Algorithm \ref{BCD} and update it according to~(\ref{penalty}), i.e., $\mathcal{F}\left(\tilde{\mathbf{r}}_{n}^{\left(t\right)}\right)$.
		\IF{$\mathcal{F}\left(\tilde{\mathbf{r}}_{n}^{\left(t\right)}\right)>\mathcal{F}\left (\tilde{\mathbf{r}}_{n,pbest}\right)$}
		\STATE Update $\tilde{\mathbf{r}}_{n,pbest}\leftarrow \tilde{\mathbf{r}}_{n}^{(t)}$.
		\ENDIF
		\IF{$\mathcal{F}\left(\tilde{\mathbf{r}}_{n}^{\left(t\right)}\right)>\mathcal{F}\left (\tilde{\mathbf{r}}_{gbest}\right)$}
		\STATE Update $\tilde{\mathbf{r}}_{gbest}\leftarrow \tilde{\mathbf{r}}_{n}^{(t)}$.
		\ENDIF
		\ENDFOR
		\ENDFOR
		\STATE Obtain the suboptimal APV $\tilde{\mathbf{r}}=\tilde{\mathbf{r}}_{gbest}$.
		\STATE Calculate the corresponding receive combining matrix $\mathbf{W}$ \quad and transmit power matrix $\mathbf{P}$ according to Algorithm~\ref{BCD}. 
		\RETURN $\tilde{\mathbf r},\mathbf{W}, \mathbf{P}$.
	\end{algorithmic}
\end{algorithm}

\section{Simulation Results}\label{sec_SimulationResults}
In the simulations, we consider a scenario where $K=12$ FPA-users are served by the BS equipped with $M=16$ MAs, and the distance between user $k$ and the BS is
assumed to be a random variable following uniform distributions, i.e., $d_k \sim \mathcal{U}[20,100], 1\le k \le 12$. The carrier wavelength is set as $\lambda=0.1$ meter (m) and the moving region of MAs at the BS is set as a square area of size $[-3\lambda/2,3\lambda/2]\times [-3\lambda/2,3\lambda/2]$. We adopt a geometry channel model, in which the numbers of receive paths for all users are the same, i.e., $L_k=L=10, 1\le k \le K$. For each user, each element of the PRV is an i.i.d. CSCG random varible, i.e., $g_{k,l} \sim \mathcal{CN}(0,\rho d_k^{-\alpha}/L), 1\le k \le K, 1 \le l \le L$, where $\rho d_k^{-\alpha}$ is the expected channel gain of user $k$ with $\rho= -40 \text{ dB}$ representing the path loss at the reference distance of 1 m, and $\alpha=2.8$ denoting the path loss exponent. 
The maximum transmit power and noise power are set as $p_\mathrm{max}= 10 \text{ dBm}$ and $\sigma^2 = -80 \text{ dBm}$, respectively.
The elevation and azimuth AoAs for each user are assumed to be i.i.d. variables following the uniform distribution over $[-\pi/2,\pi/2]$, i.e., $\theta_{k,l},\phi_{k,l} \sim \mathcal{U}[-\pi/2,\pi/2], 1\le k \le K, 1 \le l \le L$. 
The parameters in Algorithm \ref{PSO} are set as $N=200$, $T=300$, $c_1=c_2=1.4$, $\tau=10$, $\xi=\epsilon=10^{-3}$ and $\omega$ linearly decreases from 0.9 to 0.4 during the iterations, respectively. Each curve in the simulation figures is the average result over $10^3$ user distributions and channel realizations. Two benchmark schemes for comparison with our proposed APV optimization are defined as follows. For the FPA scheme, it is assumed that BS is equipped with FPA-based uniform planar array, spaced by $\lambda/2$. For the alternating position selection (APS) scheme, the receive moving region is quantized into discrete locations with equal-distance $\lambda/2$ and each MA's position is alternately selected with the others being fixed. 

\begin{figure}[t]
	\begin{center}
		\includegraphics[width=6.5 cm]{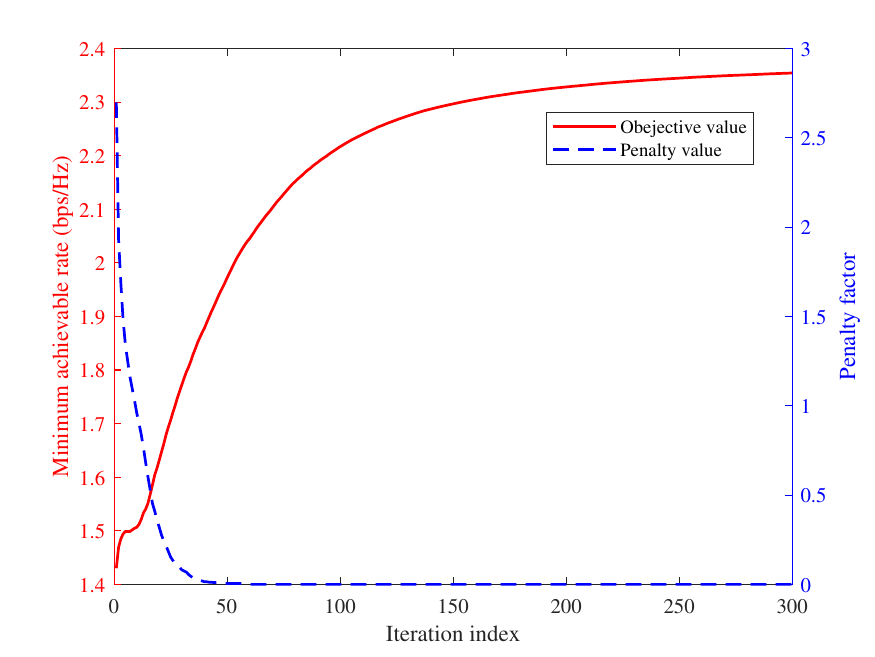}
		\caption{Objective value and penalty value versus iteration index for the proposed Algorithm \ref{PSO}.}
		\label{fig:OvsP}
	\end{center}
	\vspace{-0.7cm}
\end{figure}

First, in Fig.~\ref{fig:OvsP}, the convergence of the proposed algorithms for the MA-aided multiuser communication system is presented. Moreover, in order to validate the effectiveness of our proposed adaptive penalty factor in~(\ref{penalty}), we illustrate the penalty value versus the iteration index.
As can be observed, the minimum achievable rate of all users increases with the iteration index and remains nearly unchanged after 250 iterations, which demonstrates fast convergence performance.
In addition, the penalty value remains zero after 50 iterations, which guarantees that minimum inter-MA distance is satisfied. Particularly, the minimum achievable rate of all users increases from 1.44 bps/Hz to 2.36 bps/Hz, which yields about $63\%$ performance improvement.

\begin{figure}[t]
	\begin{center}
		\includegraphics[width=6.5 cm]{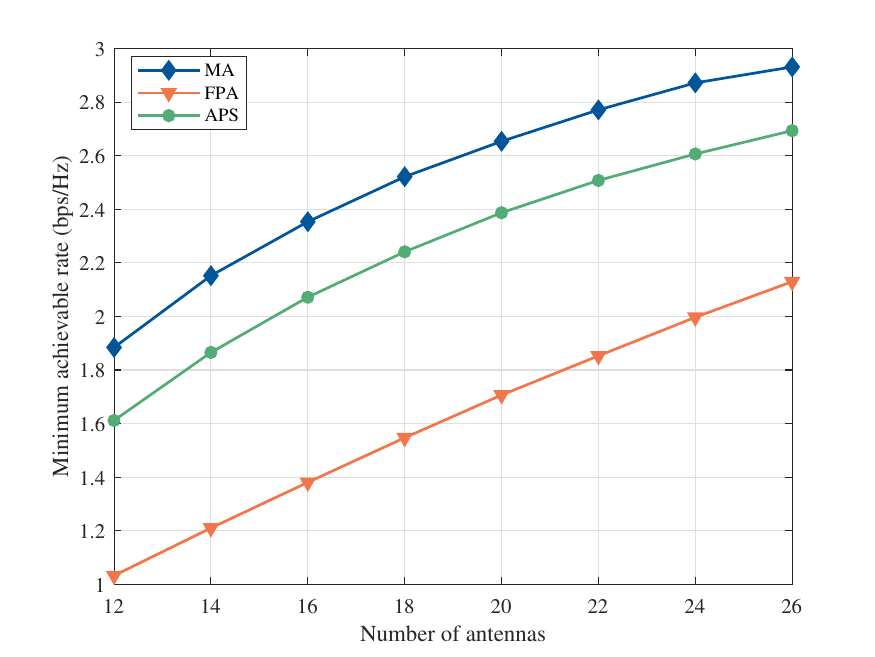}
		\caption{Minimum achievable rate for different schemes versus number of antennas.}
		\label{fig:RvsM}
	\end{center}
\vspace{-0.2cm}
\end{figure}

\begin{figure}[t]
	\begin{center}
		\includegraphics[width=6.5 cm]{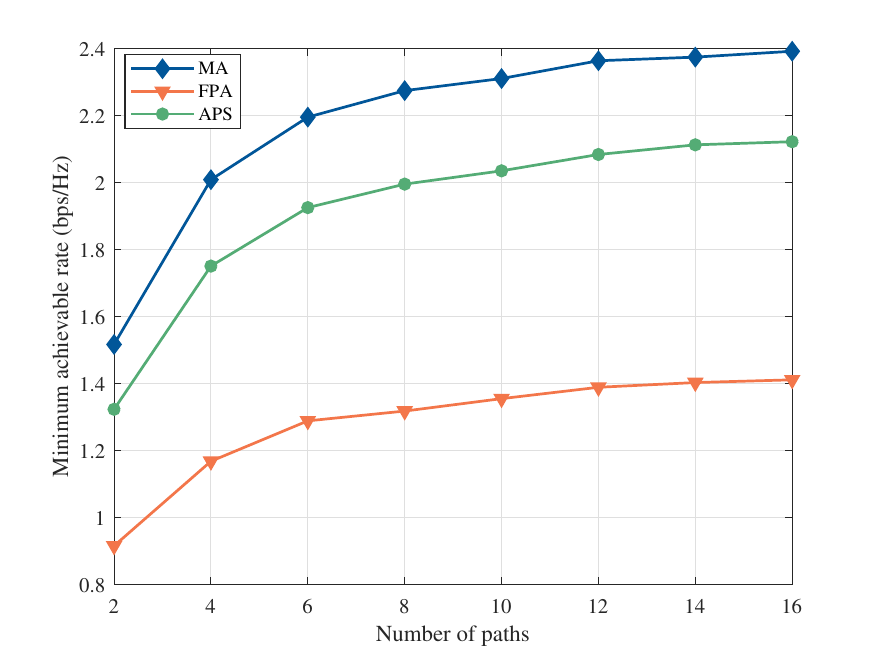}
		\caption{Minimum achievable rate for different schemes versus number of paths.}
		\label{fig:RvsL}
	\end{center}
\vspace{-0.2cm}
\end{figure}

Next, in Fig~\ref{fig:RvsM}. we compare the minimum achievable rates for different schemes versus the number of antennas. It can be seen that the proposed scheme outperforms all other benchmark schemes. With the increasing number of antennas, the minimum achievable rate increases because there is an improvement in spatial diversity gain  and beamforming gain. Compared to the conventional FPA scheme, the proposed MA scheme can leverage spatial freedom to significantly reduce the number of antennas required for the same rate performance. 

Finally, in Fig. \ref{fig:RvsL}, we compare the minimum achievable rates for different schemes versus the number of channel paths. The minimum achievable rate increases with the number of channel paths for all schemes since more paths lead to higher spatial diversity gain and lower correlation among the channel vectors for multiple users. Besides, the MA scheme can make use of the channel variation to further reduce the channel correlation. Thus, the increasing rate of MA scheme is much higher than that of FPA scheme. In addition, it is shown that the minimum achievable rate stays at a very large value and is almost unchanged when the number of paths is large enough, i.e., 14 or 16. This is because a larger region is required for antenna moving to fully exploit more spatial diversity with increasing number of channel paths. 

\section{Conclusion}
In this paper, we proposed a new BS architecture with multiple MAs to improve the multiuser communication rate performance as compared to traditional BS mounted with FPAs. We first model the multiuser channel as a function of the APV to characterize the multi-path response between the multiple MAs at the BS and the single FPA at each user. Then, based on this channel model, a joint optimization problem was formulated for designing the MA positioning, receive combining, and transmit power control to maximize the minimum achievable rate among multiple users, under the constraints of finite moving region of MAs, minimum inter-MA distance, and maximum transmit power of each user. To solve this non-convex optimization problem with highly-coupled variables, we developed a two-loop iterative algorithm based on PSO. Simulation results demonstrated that compared to FPA-based systems, our proposed solution for MA-aided uplink multiuser communication systems can significantly improve the rate performance by exploiting the new design DoF via antenna position optimization.

\bibliographystyle{IEEEtran} 
\bibliography{IEEEabrv,manuscript_MA_conf}

\begin{thebibliography}{10}
\providecommand{\url}[1]{#1}
\csname url@samestyle\endcsname
\providecommand{\newblock}{\relax}
\providecommand{\bibinfo}[2]{#2}
\providecommand{\BIBentrySTDinterwordspacing}{\spaceskip=0pt\relax}
\providecommand{\BIBentryALTinterwordstretchfactor}{4}
\providecommand{\BIBentryALTinterwordspacing}{\spaceskip=\fontdimen2\font plus
\BIBentryALTinterwordstretchfactor\fontdimen3\font minus
  \fontdimen4\font\relax}
\providecommand{\BIBforeignlanguage}[2]{{%
\expandafter\ifx\csname l@#1\endcsname\relax
\typeout{** WARNING: IEEEtran.bst: No hyphenation pattern has been}%
\typeout{** loaded for the language `#1'. Using the pattern for}%
\typeout{** the default language instead.}%
\else
\language=\csname l@#1\endcsname
\fi
#2}}
\providecommand{\BIBdecl}{\relax}
\BIBdecl

\bibitem{Ma20216G}
M.~Matthaiou, O.~Yurduseven, H.~Q. Ngo, D.~Morales-Jimenez, S.~L. Cotton, and
  V.~F. Fusco, ``The road to {6G}: Ten physical layer challenges for
  communications engineers,'' \emph{{IEEE} Commun. Mag.}, vol.~59, no.~1, pp.
  64--69, Jan. 2021.

\bibitem{Wang20236G}
C.-X. Wang, X.~You, X.~Gao, X.~Zhu, Z.~Li, C.~Zhang, H.~Wang, Y.~Huang,
  Y.~Chen, H.~Haas, J.~S. Thompson, E.~G. Larsson, M.~D. Renzo, W.~Tong,
  P.~Zhu, X.~Shen, H.~V. Poor, and L.~Hanzo, ``On the road to {6G}: Visions,
  requirements, key technologies, and testbeds,'' \emph{IEEE Commun. Surveys
  Tuts.}, vol.~25, no.~2, pp. 905--974, Secondquarter 2023.

\bibitem{HBF2017MA}
A.~F. Molisch, V.~V. Ratnam, S.~Han, Z.~Li, S.~L.~H. Nguyen, L.~Li, and
  K.~Haneda, ``Hybrid beamforming for massive {{MIMO}}: A survey,''
  \emph{{IEEE} Commun. Mag.}, vol.~55, no.~9, pp. 134--141, Sep. 2017.

\bibitem{MIMO2021LQ}
Q.~Li, X.~Yu, M.~Xie, N.~Li, and X.~Dang, ``Performance analysis of uplink
  massive spatial modulation {MIMO} systems in transmit-correlated rayleigh
  channels,'' \emph{China Commun.}, vol.~18, no.~2, pp. 27--39, Feb 2021.

\bibitem{zhu2023movable}
L.~Zhu, W.~Ma, and R.~Zhang, ``Movable antennas for wireless communication:
  Opportunities and challenges,'' \emph{arXiv prepint arXiv:2306.02331}, 2023.

\bibitem{zhu2022modeling}
L.~{Zhu}, W.~{Ma}, B.~{Ning}, and R.~{Zhang}, ``Modeling and performance
  analysis for movable antenna enabled wireless communications,'' \emph{arXiv
  prepint arXiv:2210.05325}, 2022.

\bibitem{ma2022mimo}
W.~Ma, L.~Zhu, and R.~Zhang, ``{{{MIMO}}} capacity characterization for movable
  antenna systems,'' \emph{arXiv prepint arXiv:2210.05396}, 2022.

\bibitem{MUMA2023ZL}
L.~{Zhu}, W.~{Ma}, B.~{Ning}, and R.~{Zhang}, ``Movable-antenna enhanced
  multiuser communication via antenna position optimization,'' \emph{arXiv
  prepint arXiv:2302.06978}, 2023.

\bibitem{ma2023compressed}
W.~Ma, L.~Zhu, and R.~Zhang, ``Compressed sensing based channel estimation for
  movable antenna communications,'' \emph{arXiv prepint arXiv:2306.04333},
  2023.

\bibitem{MMSE2022Lin}
Q.~Lin, H.~Shen, and C.~Zhao, ``Learning linear {MMSE} precoder for uplink
  massive {MIMO} systems with one-bit adcs,'' \emph{IEEE Wireless Commun.
  Lett.}, vol.~11, no.~10, pp. 2235--2239, Oct 2022.

\bibitem{Maxmin2008Nace}
D.~Nace and M.~Pioro, ``Max-min fairness and its applications to routing and
  load-balancing in communication networks: a tutorial,'' \emph{IEEE Commun.
  Surveys Tuts.}, vol.~10, no.~4, pp. 5--17, 2008.

\bibitem{PSO2019Zhu}
L.~Zhu, J.~Zhang, Z.~Xiao, X.~Cao, D.~O. Wu, and X.-G. Xia, ``Joint {Tx-Rx}
  beamforming and power allocation for {5G} millimeter-wave non-orthogonal
  multiple access networks,'' \emph{{IEEE} Trans. Commun.}, vol.~67, no.~7, pp.
  5114--5125, Jul. 2019.

\bibitem{DE2011DS}
S.~Das and P.~N. Suganthan, ``Differential evolution: A survey of the
  state-of-the-art,'' \emph{{IEEE} Trans. Evol. Comput.}, vol.~15, no.~1, pp.
  4--31, Feb 2011.

\end{thebibliography}

\end{document}